# New data on the systematics and interrelationships of sawfishes (Elasmobranchii, Batoidea, Pristiformes)

P. P. Deynat

Muséum National d'Histoire Naturelle, Département Milieux et peuplements
aquatiques USM 0403 « biodiversité et dynamique des communautés aquatiques »,
43 rue Cuvier, 75231 Paris cedex 05, France

## INTRODUCTION

The order Pristiformes (sawfishes) (Compagno, 1973), contains seven species in two genera (Compagno, 1999). These cartilaginous, benthic fishes present a worldwide distribution in tropical and subtropical coastal waters of Atlantic and Indo-Pacific areas, including estuaries and rivers mouths, and many studies have been conducted on their morphology and ecology (Garman, 1913; Bigelow & Schroeder, 1953; Last & Stevens, 1994; Compagno & Cook, 1995; Zorzi, 1995; Compagno & Last, 1999).

Pristiformes, with the single family Pristidae, are divided into two genera, Pristis Linck 1790 and Anoxypristis White & Moy-Thomas, 1941. Pristis comprises six putative species: Pristis pectinata Latham, 1794, Pristis pristis (L., 1758), Pristis zijsron Bleeker, 1851, Pristis microdon Latham, 1794, Pristis perotteti Müller & Henle, 1841 and Pristis clavata Garman, 1906. The genus Anoxypristis is represented by a single Indo-Pacific species, Anoxypristis cuspidata (Compagno, 1999).

Interrelationships among the species have never been clearly defined and the status of some species is still uncertain because of the lack of comparative material (Stehmann, 1990; Ishihara et al., 1991; Taniuchi & Shimizu, 1991). As part of the ongoing Odontobase$_1$ project (which is an attempt to identify chondrichtyans based on their isolated dermal elements; Deynat, 1996, 2000, 2003a, b), the dermal denticles of sawfishes were studied in detail. The aim of this paper is to show that the dermal denticles of sawfishes are useful in elucidating their systematics and in providing new information regarding the supposed synonymies of some nominal species of Pristis.

## MATERIALS AND METHODS

### MEASUREMENTS

Measurements of the specimens were made according to Bigelow & Schroeder (1953). Total length ($L_T$) is taken from the tip of the snout to the distal point of the dorsal lobe of the caudal fin. Length of the isolated 'saw' is measured from the tip to the base of the saw.

### TERMINOLOGY

Terminology of the dermal denticles and preparation of the skin samples follows Deynat & Séret (1996). Other parts of the body, including the fins and the buccopharyngeal cavity, have been examined in order to compare the variations of shape and arrangement of the denticles.

### SAMPLE COLLECTIONS

Morphological comparative studies both at the intraspecific and interspecific levels have been made on the specimens indicated in the Appendix. Institutional acronyms follow Leviton et al. (1985).

## RESULTS

Body, and in most cases, the buccopharyngeal cavity, is covered with numerous closely set dermal denticles variably arranged within both Pristis and Anoxypristis. Pristidae are devoid of any tubercles, thorns and bucklers (Garman, 1913).

ANOXYPRISTIS

In Anoxypristis, the body is partially covered by dermal denticles irregularly distributed on the 'saw' and mainly located on the anterior edges of the head and fins. The skin of the foetuses and younger neonates up to 642mm $L_T$ is completely naked. The first denticles appear in a 710mm $L_T$ neonate (MNHN A7909), on the anterior part of both dorsal and ventral sides of the saw, on the anterior edge of all fins and, sparsely, in the interorbital area of larger specimens [Fig. 1(a), (b)].

Denticles on both sides of the saw present a similar shape over the entire length of the saw. They are deeply imbricated and show an ovoid to polygonal crown, supported by a short and well differentiated peduncle, reposing on a star-shaped basal plate. Lateral edges of the saw present larger denticles to the level of insertion of the rostral teeth.
Denticles on the anterior edges of fins are similar in shape and arrangement to the denticles of the saw and can only be distinguished by their larger size. Study of different specimens of Anoxypristis shows that the body is completely devoid of any tubercles, and buccopharyngeal denticles are missing in all developmental stages. One of the larger specimens examined (MNHN 3459, female 785mm $L_T$) shows some sparsely distributed denticles in the interorbital and anterorbital areas, without a well defined arrangement. These denticles have a flat crown, slightly erected at its posterior tip, without any distinctive relief. The crown presents three posterior tips: one median tip and two lateral [Fig. 1(a), (b)]. In adults, the dermal covering shows widely spaced denticles.

PRISTIS

In Pristis, the dermal denticles on both sides of the body and tail are regularly imbricated. The length of the crown ranges from 100 to 150 mm. The starshaped basal plate is surrounded by a short and feebly differentiated peduncle, supporting a leaf-shaped crown, slightly erected at its posterior side and marked by a more or less differentiated posterior tip [Figs 1(c) - (f) and 2(e), (f)].
The study of different specimens shows that a variation of the shape of the crown separates the genus Pristis into two groups. The first group, comprising P. pristis, P. perotteti and P. microdon is characterized by the development of crown ridges and furrows on the anterior part of the denticles in all developmental stages. Denticles of the dorsal side of the body show a leaf-shaped crown, with three to five ridges extending from the anterior edge of the crown to the mid-length of the denticle [Fig. 1(e), (f)]. These ridges are developed on each denticle of the dorsal and ventral sides of the body, tail and fins. In larger specimens, ridges are less marked and three to nine furrows are present. The smallest denticles present only one median ridge. During replacement, the next denticles show in addition a pair of parasagittal ridges; in larger specimens, there are deep furrows between these ridges [Fig. 1(e), (f)]. Dermal denticles of the ventral side of the body present a very similar crown with furrows on the first anterior third of the denticle [Fig. 2(e)].
The second group, comprising P. pectinata, P. clavata and P. zijsron, is characterized by dermal denticles with a flat and oval-shaped or rounded crown devoid of any relief in all development stages. Denticles of the dorsal part of the body show a similar morphology and present a leaf-shaped crown terminating at a feeble tip, although the tip is more pronounced in adults (Garman, 1913) [Fig. 1(c), (d)]. The study of the morphological variations of these denticles shows a differentiation of the crown, becoming more elongate and convex in the posterior part of the body. Denticles of the ventral side are smaller and set more in a pavement-like pattern (Bigelow & Schroeder, 1953) [Fig. 2(f)].
Within the two groups discussed above, denticles of both dorsal and ventral sides of the body generally show similar morphology (dorsoventral isomorphy sensu Deynat & Se´ ret, 1996), but with the denticles of the ventral side being smaller (Bigelow & Schroeder, 1953). Furthermore, they can be differentiated by the presence or absence of crown ridges developed on the anterior part of the denticles.

DERMAL COVERING OF THE SAW

As for A. cuspidata, dermal denticles of the species of the genus Pristis present a similar morphology on all or part of both sides of the saw. They can be differentiated by their larger size on the lateral edges of the saw at the contact zone between the 'teeth' and the 'saw'. The crowns of the denticles are deeply imbricated, creating a plain surface without any relief [Fig. 2(a), (b)]. This local differentiation of polygonal denticles can also be found on the anterior edge of the fins and snout where the denticles do not show any ridge, nor any distinct posterior tip. This character is also found in Anoxypristis, and is frequently observed among sharks. It could be a primitive character linked to ecomorphological variations and conserved within Pristiformes (Wallace, 1967).
Rostral denticles seem to present a similar shape all over the length of the saw, but a detailed examination shows that a progressive antero-posterior modification of the crown morphology occurs among species in which crown ridges are present (P. pristis, P. perotteti and P. microdon). Differentiation of the crown ridges occurs on the posterior third of the saw to the tail: denticles are devoid of any relief on the anterior two-thirds of the length of the saw; ridges are noticeably differentiated on denticles c. 20 cm from the anterior edge of the eyes [Fig. 2(a), (b)].

MODIFIED DENTICLES

As noted above, species of Pristis have an additional dermal covering of large, deeply imbricated and polygonal denticles. These peculiar denticles are located on the anterior edge of the fins, lateral parts of the saw, and along the caudal keels, as indicated in P. pectinata by Garman (1913).
All specimens studied had small denticles in the buccopharyngeal cavity [Fig. 2(c), (d)]. These denticles have a thick and leaf-shaped crown, with three to five strong ridges. The crown is elongated and slender in P. pristis and P. zijsron, but more enlarged in P. pectinata.

DISCUSSION

Previous studies of the characteristics of the dermal covering of sawfishes have shown that a separation can be made between the genera Anoxypristis and Pristis on the one hand, and among the different species of Pristis in two distinct groups on the other (Deynat, 1996, 2000).
Pristiformes share a typical dermal covering characterized by: (1) large denticles with a polygonal crown which are deeply imbricated and arranged as a mosaic on the anterior edges of the fins and over most of the length of the saw; (2) denticles of both sides of the body comprising a leaf-shaped crown on a welldifferentiated peduncle and basal plate; (3) absence of tubercles or hypertrophied denticles.
The main differences between the dermal covering of Anoxypristis compared to Pristis are in the mode of covering of both dorsal and ventral parts of the body, the presence of buccopharyngeal denticles, and the comparative morphology of the crown of the denticles. The covering of both the dorsal and ventral parts of the body and tail at an equivalent growth stage constitutes a criterion of intergeneric taxonomic differentiation among Pristiformes.
Anoxypristis is characterized as naked (Fowler, 1941), or more or less covered with denticles in adults (Compagno, 1973; Taniuchi et al., 1991; Last & Stevens, 1994). Differentiation of larger denticles on the anterior edges of the fins, and the ontogenic differentiation of denticles on both dorsal and ventral sides of the saw in juveniles of Anoxypristis, indicate a typical postnatal development of the
dermal covering.
A dorsal keel comprising differentiated denticles (large and polygonal or heart-shaped) has been described in P. zijsron (Fowler, 1941; Qureshi, 1972) and in Anoxypristis (Annandale, 1909), but has only been observed in one juvenile specimen of P. pectinata (pers. obs.). This mid-dorsal keel, comprising larger denticles of the same kind as the body denticles, extends from just behind the spiracles to the base of the second dorsal fin. The keel is differentiated at the level of the posterior tip of the free edge of the first dorsal fin. Furthermore, Annandale (1909) noticed a dark line of flat and heart-shaped denticles, from the area just before the first dorsal fin to the level of the anterior edge of the pelvics. A dark mid-dorsal line appears in foetuses of P. zijsron and P. pectinata, from the posterior level of the eyes to the level of the second dorsal fin (MNHN 1967-948), but this embryonic line does not correspond to the keel of modified denticles, as observed in the juvenile specimen of P. pectinata. This dark dorsal line seems to result from a transitory differentiation of the dermal covering in juvenile stages of P. zijsron and P. pectinata, and is not considered to be of special taxonomic importance. Presence of a well differentiated mid-dorsal keel as noted by Annandale (1909) has never been recorded in adults of P. pectinata and P. zijsron in which denticles of the body are of a similar size.

Morphology of the denticles of the body distinguishes Anoxypristis and Pristis because only Anoxypristis presents tricuspidate denticles rather than monocuspidate ones. In Anoxypristis, the morphology and arrangement of the denticles on the ventral side of the body and tail are mostly unknown and it is therefore difficult to make a reliable comparison between both genera. Buccopharyngeal denticles are always present in Pristis, but their absence in the juvenile specimens of Anoxypristis available for this study perharps does not constitute a reliable criterion for differentiation, without knowing the character state in adults. In Pristis, buccopharyngeal denticles are always differentiated and present a similar morphology in all species, independent of the morphology of the denticles of both dorsal and ventral parts of the body and tail [Fig. 2(c), (d)].

Pristis shows an early and homogenous covering of both sides of the body and tail beginning in neonates, except for the lateral edges of the saw where the covering appears during later growth and is linked to the development of the rostral teeth. Among the different species of Pristis examined, the morphology of the denticles never changes from the early stages of development, and the presence or absence of the anterior crown ridges on the denticles of both sides of the body and tail is a reliable taxonomic character for distinguishing groups of species. Based on their denticles, Pristis can be divided into two groups, the first one comprising P. pristis, P. perotteti and P. microdon, and the second one comprising P. pectinata, P. zijsron and P. clavata. The first group is characterized by denticles, on both the dorsal and ventral sides of the body, with leafshaped crowns and well developed anterior keels, whereas the second group is characterized by smooth ovoid denticles without any superficial relief on both dorsal and ventral parts of the body and tail. Features of the dermal covering in Pristiformes thus provide useful systematic data, even if it is impossible to separate the species within the two groups of Pristis.

Traditionally, there have been many problems associated with the taxonomy of sawfishes. The systematics and phylogenetic interrelationships of sawfishes have not been clearly elucidated (Ishihara et al., 1991; Taniuchi & Shimizu, 1991) for several reasons. Specimens are difficult to obtain, especially large ones, because some species are threatened and need to be protected. It has also been difficult to obtain growth series for analysis of valuable ontogenic characters (Compagno & Cook, 1995).

Several morphological data have been used to distinguish sawfishes. Duméril (1865) was the first to use the morphology of the rostral teeth to separate 11 species of Pristis. Hoffman (1912) first distinguished P. cuspidatus from other species of Pristis by morphological characteristics, and White & Moy-Thomas (1941) created the genus Anoxypristis for this species. Garman (1913), and subsequently Bigelow & Schroeder (1953), used the relative position of the fins and the number of rostral teeth to recognize six species of Pristis. These characters have been used mainly to separate the extant species of Pristis. If only denticle morphology and arrangement are considered, separation of A. cuspidatus from the genus Pristis is clearly supported by the present work, agreeing with Hoffman's (1912) results.

Within the genus Pristis, the interpretation of dermal denticles is evidently more complex, because several misidentifications have been reported (Last & Stevens, 1994; Zorzi, 1995; Eschmeyer, 1998). Results of the present study of the dermal covering do not allow a diagnosis of the different species belonging to Pristis to be made. In creating two distinct groups, however, the data can be related to the interrelationships of Pristidae suggested by several authors using other morphological data (Bigelow & Schroeder, 1953; Kailola, 1987; Zorzi, 1995).

The identification of two groups coincides with the work of Zorzi (1995) and Compagno & Cook

(1995) who indicated that two groups of Pristis could be identified using features of external morphology (such as the shape of the saw and body, presence or absence of a caudal lobe, and number of rostral teeth; Zorzi, 1995). The first group (the P. pristis group), comprising P. pristis, P. microdon and P. perotteti, is characterized by a broadly tapering robust saw, 15-22 rostral teeth and a thick body, but the relationships between the different species have never been clearly elucidated (Zorzi, 1995). The second group (the P. pectinata group), comprising P. pectinata, P. zijsron and P. clavata, is characterized by a more slender saw and body, and 23-35 pairs of rostral teeth (Zorzi, 1995).
Pristis clavata and P. pristis, considered by some authors to be very similar, have been placed in synonymy by Paxton et al. (1989). Characteristics of their dermal covering, however, show that P. clavata is more closely related to P. pectinata than to P. pristis.

KEY FOR DERMAL COVERING FEATURES AMONG PRISTIFORMES

1a. Dermal covering constituted by close-set monocuspidate dermal denticles completely covering both sides of the body, fins and buccopharyngeal cavity . . . . . . . . . 2
1b. Dermal covering constituted by tricuspidate dermal denticles sparsely covering both sides of the body and fins, and absent in the buccopharyngeal cavity . . . . . . . . . A. cuspidata
2a. Denticles of both dorsal and ventral parts of the body with well marked ridges and furrows on their anterior edge. . . . . . . . P. pristis, P. perotteti, P. microdon
2b. Denticles of both dorsal and ventral parts of the body devoid of any ridges and furrows on their anterior edge. . . . . . . . P. pectinata, P. zijsron, P. clavata

I thank P. Last for the loan of material, M. Stiassny, M. Carvalho, S. Raredon and J. Williams for their welcome during my stays in New York and Washington, D. Guillaumin for the SEM photographs, A. Janoo for reviewing the English version of this manuscript, P. Pruvost and J. Gregorio for assistance during this study and an unknown referee from providing constructive remarks. Special thanks to E. Baldzinger for providing references and V. Faria for interesting suggestions. The 'Fondation Bleustein-Blanchet pour la Vocation' attributed a grant to PPD for the Odontobase$_1$ project. This paper is dedicated to the memory of G. Dingerkus.

References

Annandale, N. (1909). Report on the fishes taken by the Bengal Fisheries Steamer ''Golden Crown''. Part I, Batoidei. Memoirs of the Indian Museum, Calcutta 2, 1-58.
Bigelow, H. B. & Schroeder, W. C. (1953). Sawfishes, guitarfishes, skates and rays; Chimaeroids. In Fishes of the Western North Atlantic, No. 1 (Tee-Van, J., Breder, C. M., Pair, A. E., Schroeder, W. C. & Schultz, L. P., eds), pp. 1-514. New Haven, CT: Memoirs of the Sears Foundation for Marine Research, Yale University.
Compagno, L. J. V. (1973). Interrelationships of the living elasmobranchs. In Interrelationships of Fishes, Suppl. 1 (Greenwood, P. H., Miles, R. S. & Patterson, C., eds), pp. 15-61. New York: Academic Press.
Compagno, L. J. V. (1999). Checklist of living Elasmobranchs. In Sharks, Skates, and Rays, the Biology of Elasmobranch Fishes (Hamlett, W. C., ed.), pp. 471-498. Baltimore, MD: John Hopkins University Press.
Compagno, L. J. V. & Cook, S. F. (1995). The exploitation and conservation of freshwater elasmobranchs: status of taxa and prospects for the future. In The Biology of Freshwater Elasmobranchs, a Symposium to Honor Thomas B. Thorson (Oetinger, M. I. & Zorzi, G. D., eds). Journal of Aquariculture & Aquatic Sciences VII, 62-90.
Compagno, L. J. V. & Last, P. R. (1999). Pristidae. Sawfishes. In FAO Identification Guide for Fishery Purposes. The Living Marine Resources of the Western Central Pacific (Carpenter, K. E. & Niem, V., eds), pp. 1410-1417. Rome: FAO.
Deynat, P. (1996). Applications de l'étude du revêtement cutané des Chondrichtyens à la systématique phylogénétique des Pristiformes et Rajiformes sensu Compagno, 1973 (Elasmobranchii, Batoidea). PhD thesis, University of Paris VII Denis Diderot- Museum national d'histoire Naturelle.
Deynat, P. (2000). Dermal denticle morphology within Batoid rays: a review. In Proceedings of the 3rd Meeting of the European Elasmobranch Association, Boulognesur- Mer, 1999 (Séret, B. & Sire, J.-Y., eds), pp. 15-27. Paris: Société française d'Ichtyologie & IRD.
Deynat, P. (2003a). Unusual polarity pattern of dermal denticles in the whale shark Rhincodon typus (Elasmobranchii, Orectolobiformes, Rhincodontidae). Biociências 11, 61-64.

Deynat, P. (2003b). Les requins à fleur de peau. Apnéa 153, 12-13.
Deynat, P. & Séret, B. (1996). Le revêtement cutané des raies (Chondrichthyes, Elasmobranchii, Batoidea). I: Morphologie et arrangement des denticules cutanés. Annales des Sciences Naturelles, Zoologie 17, 65-83.
Duméril, A. H. A. (1865). Histoire naturelle des poissons ou ichtyologie générale. Tome premier. Elasmobranches, plagiostomes et holocephales, ou chimères. Première partie. Paris: Librairie encyclopédique de Roret.
Eschmeyer, W. N. (Ed.) (1998). Catalog of Fishes. Special Publication, (3 vols). San Francisco, CA: California Academy of Sciences.
Fowler, H. W. (1941). Contributions to the biology of the Philippine archipelago and adjacent regions. The fishes of the groups Elasmobranchii, Holocephali, Isospondyli and Ostariophysi obtained by the United States Bureau of Fisheries Steamer "Albatross" in 1907, chiefly in the Phillipine Islands and adjacent seas. Bulletin of
the United States National Museum 100, 1-879.
Garman, S. W. (1913). The Plagiostomia (sharks, skates and rays). Memoirs of the Museum of Comparative Zoology 36, 1-528.
Hoffman, L. (1912). Zur kennis des neurocraniums des Pristiden und Pristiophoriden. Zoologische Jahrbuecher, Anatomisch (Jena) 33, 239-360.
Ishihara, H., Taniuchi, T., Sano, M. & Last, P. R. (1991). Record of Pristis clavata Garman from the Pentecost River, Western Australia, with Brief Notes on its osmoregulation, and comments on the systematics of the Pristidae. In Studies on Elasmobranchs Collected from Seven River Systems in Northern Australia and Papua New Guinea (Shimizu, M. & Taniuchi, T., eds), pp. 45-53. Tokyo: University Museum, University of Tokyo, Nature and Culture.
Kailola, P. J. (1987). The fishes of Papua New Guinea: A revised and annotated checklist. Vol. I: Myxinidae to Symbranchidae. Research Bulletin, N_ 41, Port Moresby: Department of Fisheries and Marine Resources. Last, P. R. & Stevens, J. D. (1994). Sharks and Rays of Australia. Australia: CSIRO.
Leviton, A. E., Gibbs Jr, R. H., Heal, E. & Dawson, C. E. (1985). Standards in herpetology and ichthyology: Part I. Standard symbolic codes for institutional resource collections in herpetology and ichthyology. Copeia 1985, 802-832.
Paxton, J. R., Hoese, D. F., Allen, G. R. & Hanley, J. E. (1989). Pisces. Petromyzontidae to Carangidae. Zoological Catalogue of Australia, Vol. 7. Canberra: Australian Government Publishing Service.
Qureshi, M. R. (1972). Sharks, skates and rays of the Arabian Sea. Pakistan Journal of Scientific and Industrial Research 15, 294-311.
Stehmann, M. (1990). Pristidae. In Checklist of the Fishes of the Eastern Tropical Atlantic (CLOFETA), Vol. 1 (Quero, J. C., Hureau, J. C., Karrer, C., Post, A. & Saldanha, L., eds), pp. 51-54. Lisbon: JNICT, Paris: SEI & Paris: UNESCO.
Taniuchi, T. & Shimizu, M. (1991). Elasmobranchs collected from seven river systems in northern Australia and Papua New Guinea. In Studies on Elasmobranchs Collected from Seven River Systems in Northern Australia and Papua New Guinea (Shimizu, M. & Taniuchi, T., eds), pp. 3-10. Tokyo: University Museum, University of Tokyo, Nature and Culture.
Taniuchi, T., Kan, T. T., Tanaka, S. & Otake, T. (1991). Collection and measurement data and diagnostic characters of Elsmobranchs collected from three river systems in Papua New Guinea. In Studies on Elasmobranchs Collected from Seven River Systems in Northern Australia and Papua New Guinea (Shimizu, M. & Taniuchi, T., eds), pp. 27-41. Tokyo: University Museum, University of Tokyo, Nature and
Culture.
Wallace, J. H. (1967). The Batoid Fishes of the East Coast of Southern Africa. Part I: Sawfishes and Guitarfishes. Investigational Report. Oceanographic Research Institute 15, 1-32.
White, E. I. & Moy-Thomas, J. A. (1941). Notes on the nomenclature of fossil fishes. Part II. Homonyms D-L. Annals of the Magazine of Natural History 6, 98-103.
Zorzi, G. D. (1995). The biology of freshwater elasmobranchs: an historical perspective. In The Biology of Freshwater Elasmobranchs, a Symposium to Honor Thomas B. Thorson (Oetinger, M. I. & Zorzi, G. D., eds), pp. 10-31. Journal of Aquariculture & Aquatic Sciences VII, 10-31.

APPENDIX

Specimens examined

*Anoxypristis cuspidata*: AMNH 3268, (two female foetuses of 345mm and 350mm $L_T$), Ceylon, Indian Ocean; MNHN 1236, (juvenile male, 640mm $L_T$), Malabar, Indian Ocean, Dussumier, identified by G. Dingerkus, 1982; MNHN 1234, (two juvenile specimens: one female, 642mm $L_T$ and one male, 660mm $L_T$), Pondichery, Indian Ocean, Bellanger; MNHN A7909, (juvenile female, 710mm $L_T$), Malabar, Indian Ocean, Belanger, determined by G. Dingerkus, 1982; MNHN 1250, (juvenile male,

810mm L$_T$), Malabar, Indian Ocean, Dussumier, determined by G. Dingerkus, 1982, neotype of P. cuspidatus; MNHN 3459, (juvenile female, 785mm L$_T$), Indian Ocean, determined by G. Dingerkus, 1982; MNHN 1986-1076, (isolated saw, 500mm L$_T$), unknown origin, determined by P. Deynat, 1994; MNHN 1986-1075 (isolated saw, 525mm L$_T$), unknown origin, determined by P. Deynat, 1994; MNHN 1986- 1078, (isolated saw, 765mm L$_T$), Indian Ocean, Dussumier.

Pristis clavata: CSIRO-H 2757-01, (two samples of skin, female, 810mm L$_T$), Tasmania.

Pristis microdon: CSIRO-H 2755-01, (two samples of skin, female, 1080mm L$_T$), Tasmania; MNHN 1994-647 (female, 1910mm L$_T$), unknown origin, determined by P. Deynat, 1994; MNHN A9699, (juvenile female, 2830mm L$_T$), Senegal, Perottet, determined by J. D. McEachran & B. Se´ ret, 1986; MNHN 1994-646, (male, 3400mm L$_T$), unknown origin, determined by P. Deynat, 1994; MNHN 1902-255, (female, 5100mm L$_T$), Ivory Coast, Blanc.

Pristis pectinata: MNHN 1251, (two females fetuses, 220 and 225mm L$_T$), Gabon, Aubry-Lecomte, determined by G. Dingerkus, 1982; MNHN 1967-949, (female, 710mm L$_T$), Red Sea, Towila Island, Dollfus, January 1929, determined by Fourmanoir & Postel; USNM 30678, (juvenile male, 710mm L$_T$), Gulf of Mexico; MNHN 1986-397, (juvenile female, 732mm L$_T$), Mauritania, Port-Etienne, determined by B. Se´ ret, 1986; MNHN A9476, (juvenile male, 740mm L$_T$), Antilles, Plee, holotype of P. acutirostris Dume´ril, 1865, holotype of P. pectinata following G. Dingerkus, 1982; MNHN B0549, (female, 770mm L$_T$), unknown origin, Adanson; MNHN 3485, (juvenile female, 810mm L$_T$), Red Sea, Ruppell, 1830, identified by G. Dingerkus, 1982, syntype of P. leptodon Dume´ril, 1865; MNHN 3486, (juvenile female, 910mm L$_T$), Red Sea, Botta, identified by G. Dingerkus, 1982, syntype of P. leptodon Dume´ril, 1865; MNHN 2607, (female, 798mm L$_T$), Indian Ocean, Re´ union Island, Nigou, determined by G. Dingerkus, 1982; MNHN 1990-0002, (juvenile male, 1043mm L$_T$), Guinea-Bissau, Cacheu, B. Se´ ret, September 1983, determined by P. Deynat, 1995; MNHN 1901-0492, (female, 1585mm L$_T$), Somalia, 1909; MNHN 9859, (female, 3750mm L$_T$), Haiti; MNHN 9858, (female, 3900mm L$_T$), Antilles, Haiti, Ricord; MNHN 3484, (isolated saw, 643mm L$_T$), French Guiana, Cayenne, holotype of P. megalodon Dume´ril, 1865; MNHN 2003-2614, (isolated saw, 820mm L$_T$), Congo, Pointe Noire, Molez, 1958, determined by P. Deynat, 2003; MNHN 1986-0626, (isolated saw, 840mm L$_T$), unknown origin, station ab69; MNHN 1903-427, (isolated saw, 970mm L$_T$), French Guiana, station ab68, 1900; MNHN 1888-253, (isolated saw, 1265mm L$_T$), Persian Gulf, Moet, 1888, station ab71; MNHN 1986-1077, (isolated saw, 1350mm L$_T$), unknown origin; MNHN 1888-252, (isolated saw, 1375mm L$_T$), Persian Gulf, Moet, 1888; MNHN 1986-625, (isolated saw, 1810mm L$_T$), unknown origin, station ab81.

Pristis perotteti: USNM 146543, (juvenile female, 780mm L$_T$), Guatemala.

Pristis pristis: MNHN A3527, (juvenile male, 850mm L$_T$), Cochinchina, Vietnam, Mekong, Jullien, determined by G. Dingerkus, 1982; MNHN 2003- 2611, (isolated saw, 800mm L$_T$), Democratic Republic of Congo; MNHN 2003-2612, (isolated saw, 854mm L$_T$), Democratic Republic of Congo; MNHN 2003-2613, (isolated saw, 1030mm L$_T$), Democratic Republic of Congo; MNHN 1994-309, (isolated saw, 1030mm L$_T$), unknown origin, Fournier, determined by P. Deynat, 1994; MNHN 1986-1074, (isolated saw, 1175mm L$_T$), unknown origin, identified by P. Deynat, 1994; MNHN 1986- 622, (isolated saw, 1220mm L$_T$), unknown origin, station ab77; MNHN 1983- 1345, (isolated saw, 1230mm L$_T$), unknown origin, station ab301, identified by P. Deynat, 1994; MNHN 1986-623, (isolated saw, 1285mm L$_T$), unknown origin, station ab76; MNHN 1986-621, (isolated saw, 1300mm L$_T$), unknown origin, station ab79; MNHN 1986-624, (isolated saw, 1350mm L$_T$), unknown origin, station ab78; MNHN 1938-0016, (isolated saw, 1520mm L$_T$), Singapore, Ruyters, station ab60.

Pristis zijsron: MNHN 1967-948, (female foetuse, 715mm L$_T$), Red Sea, Dollfus, January 1929, identified by G. Dingerkus, 1982; USNM 40003, (juvenile male, 872mm L$_T$), Australia; MNHN 1226, (juvenile male, 877mm L$_T$), Amboine, Indonesia, Bleeker, 1856, identified by McEachran & Se´ ret, 1986; AMNH 56530, (isolated saw, 245mm L$_T$), unknown origin; MNHN 1895- 0003, (isolated saw, 1430mm L$_T$), Gulf of Thailand, Pavie, 1895, station ab70.